\documentstyle[psfig]{mn}

\def\gta{ \lower .75ex \hbox{$\sim$} \llap{\raise .27ex \hbox{$>$}} }
\def\lta{ \lower .75ex\hbox{$\sim$} \llap{\raise .27ex \hbox{$<$}} }
\newcommand{\pasp}{PASP}
\newcommand{\aap}{A\&A}
\newcommand{\apj}{ApJ}

\newcommand{\apjl}{ApJ Letter}
\newcommand{\mnras}{MNRAS}
\newcommand{\aaps}{A\&ASS}
\newcommand{\nat}{Nature}

\title[]{The effects of a comptonizing corona on the
appearance of the reflection components in accreting black hole spectra}
\author[]
        {P.O. Petrucci$^1$, A. Merloni$^2$, A. Fabian$^2$, F. Haardt$^3$,
E. Gallo$^3$\\
        $^1$Osservatorio Astronomico di Brera, via Brera 28, 20121
Milano, Italy\\
        $^2$Institute of Astronomy,
Cambridge, UK\\
        $^3$Universit\'a dell'Insubria, Como, Italy}
\date{Accepted ?. Received ?}
\pagerange{\pageref{firstpage}--\pageref{lastpage}}
\pubyear{}

\begin{document}

\maketitle
\label{firstpage}

%--\title{The effects of a comptonizing corona on the appearance of the
%--reflection component in accreting black hole spectra}
%--\author{P.O. Petrucci \inst{1} \and A. Merloni \inst{2} \and F. Haardt
%--\and A. Fabian \inst{2} et al.}  \institute{Osservatorio Astronomico di
%--Brera, via Brera 28, 20121 Milano, Italy\\ \and Institute of Astronomy,
%--Cambridge, UK} \offprints{P.O. Petrucci}
%--\mail{petrucci@brera.mi.astro.it, petrucci@obs.ujf-grenoble.fr}
%--\date{Received ??/ accepted ??}

%--\abstract{
\begin{abstract}
We discuss the effects of a comptonizing corona on the appearance of the
reflection components, and in particular of the reflection hump, in the
X-rays spectra of accreting black holes. Indeed, in the framework of a
thermal corona model, we expect that part (or even all, depending on the
coronal covering factor) of the reflection features should cross the hot
plasma, and thus suffer Compton scattering, before being observed. We
have studied in detail the dependence of these effects on the physical
(i.e. temperature and optical depth) and geometrical (i.e. inclination
angle) parameters of the corona{ , concentrating on the slab geometry}
. Due to the smoothing and shifting towards high energies of the
comptonized reflection hump, the main effects on the emerging spectra
appear above 100 keV. We have also investigated the importance of such
effects on the interpretation of the results obtained with the standard
fitting procedures. We found that fitting Comptonization models, taking
into account comptonized reflection, by the usual cut--off power law +
uncomptonized reflection model, { may lead to an underestimation of}
the reflection normalization and { an overestimation of} the high
energy cut--off.  We discuss and illustrate the importance of these
effects by analysing recent observational results as those of the galaxy
NGC 4258. We also find that the comptonizing corona can produce {
and/or emphasize} correlations between the reflection features
characteristics (like the iron line equivalent width or the covering
fraction) and the X--ray spectral index similar to those recently
reported in the literature. We also underline the importance of these
effects when dealing with accurate spectral fitting of the X-ray
background.
%--\keywords{}} \titlerunning{} \authorrunning{} \maketitle
\end{abstract}


%%%%%%%%%%%%%%%%%%%%%%%%%%%%%%%%%%%%%%%%%%%%%%%%%%%%%%%%%%%%%%%%%%%%%%%%%%%%
\section{Introduction}
%%%%%%%%%%%%%%%%%%%%%%%%%%%%%%%%%%%%%%%%%%%%%%%%%%%%%%%%%%%%%%%%%%%%%%%%%%%%
The presence of secondary components in the spectra of Seyfert galaxies
and Galactic Black Hole Candidates (GBHC), such as an iron line at
$\sim$6.4 keV and a reflection hump between 10 and 50 keV, superimposed
on the primary X-ray continuum, is now well established observationally
\cite{nan94}. Discovered in the late eighties with the {\it GINGA}
satellite \cite{mat90,pou90}, they are signatures of reprocessing of the
primary X-ray emission in surrounding cold ($T\le 10^5 K$) and optically
thick matter accreting onto the central engine. Indeed, part ($\sim
10$\%) of the primary X-ray radiation may be Compton reflected at the gas
surface, producing the observed secondary features
\cite{bai78,lig88,geo91,mat91,mag95}. The remaining incident flux is
reprocessed in the UV/Soft X-ray band and is believed to form part of the
UV--soft X--ray excess (the so-called UV bump) generally observed in this
class of objects.\\

The reflections features are of great importance in testing theoretical
models of the high energy emission in compact objects since they give
crucial (while undirect) constraints on the geometry and on the nature of
the emitting regions.\\

Compton scattering of the soft photons emitted by the thick matter on a
population of hot thermal electrons is the best model to date for the
primary high energy emission in these objects. Indeed, the so-called
corona models, which assume radiative equilibrium between the thermal
comptonizing plasma and the underlyimg cold matter (generally an
accretion disk), can naturally account for the average properties of the
X-ray emission of black holes accretion flows \cite{haa91}. It has also
been shown that a patchy geometry, where the corona is disrupted in
localized blobs, appears to be in better agreement with the spectral
variability and $L_{UV}/L_{X}$ ratios observed for example in Seyfert
galaxies \cite{haa94,ste95}. { In the case of GBHC, like Cyg X-1, simple
slab--corona models even fail to fit the high signal to noise X--ray
spectra of these objects and more complex configurations are required
\cite{gie97}. The real geometry of the corona--disk system is thus
relatively difficult to constrain in the framework of this class of
models and the reflection features can give some important clues in this
respect.}\\

An alternative model for the high energy emission in black hole spectra
consists of an advection--dominated accretion flow (ADAF) near the
central engine. ADAFs are expected to occur in low mass accretion rate
objects, where the hot plama density is low enough to prevent a rapid
energy transfer from protons to electrons. Most of the accretion energy
is then advected into the central black hole rather than being radiated
(Narayan et al. 1998, and references therein).\\

The amount of reflection is expected to be different in these different
scenarios. For instance, in a disk--corona configuration we expect a
reflection fraction of the order of unity, since the reflected disk, as
seen by the hot corona, subtends a solid angle of 2$\pi$. On the other
hand, in the case of a truncated disk--ADAF model, the solid angle would
be smaller due to the lack of reflecting matter in the central
regions. For the same reason, we expect narrower iron K$\alpha$
fluorescent lines since relativistic effects, which are stronger in the
vicinity of the black hole, are negligible in the ADAF interpretation.\\

We thus expect the reflected spectrum to provide a strong obervational
test to discriminate between these distinct models and between different
possible geometries. However, various effects may complicate the
situation. For instance, the complex ionization pattern of the reflector
can strongly mask and/or modify the reflection features in comparison to
what we expect in the case of a simple neutral material or even
single--zone ionization model. Done \& Nayakshin \shortcite {don01} and
Ballantyne, Ross \& Fabian \shortcite {bal00} have recently shown that
using single--zone ionization models can severely underestimate the
reflection normalization if the accretion disk is highly ionized. It can
even produce an apparent correlation between the reflection normalization
and the spectral photon index similar to that reported by Zdziarski et
al. \shortcite {zdz99} in a sample of Seyfert galaxies and galactic black
hole candidates. The presence of a dynamic hot corona
\cite{rey97,bel99,mal00} will also strongly modify the reflection
spectrum due to the anisotropic illumination produced by relativistic
motions inside the corona.\\

We present here another important effect which will also modify the
appearance of the reflection components. Indeed, in a corona--disk
configuration, we expect that part (or even all, depending on the coronal
covering fraction) of the reflection features cross the comptonizing
plasma before being observed. In this case, the reflected photons are
also Compton scattered in the corona and the shape of the secondary
spectral components may be significantly modified as compared to the one
expected when no Comptonization in the corona is taken into account.\\

The effects of the hot corona in modifying the reflection component have
been already noted by some authors \cite{haa93b}. The aim of this paper
is thus to study more precisely the dependence of these effects on the
physical (temperature and optical depth) and geometrical (inclination
angle) parameters of the hot corona. We are also interested in the
consequence for spectral analysis of real data when these effects are
taken into account. { We mainly focus on the slab geometry where the
effects are expected to be large since the corona completely covers the
reflecting material.}

The paper is organized as follows. In section \ref{mod} we briefly
summarize the main characteristics of the Comptonization model we
used. We discuss the effects of a comptonizing corona on the reflection
hump shape, varying the corona optical depth and/or temperature and for
different inclination angles in section \ref{compeff}. In section
\ref{eweffect}, we focus on the iron line and on how the Comptonization
process affects the measurement of its equivalent width. We also
investigate, in section \ref{simu}, the importance of these effects on
the interpretation of the results obtained with the standard fitting
procedures. We discuss and illustrate the importance of these effects in
section \ref{obscons} by analysing recent observational results as those
of the galaxy NGC 4258. We also show that the comptonizing corona can
produce apparent correlations between the reflection features
characteristics (like the iron line equivalent width or the covering
fraction) and the X--ray spectral index similar to those recently
reported in the literature. We finaly underline the importance of these
effects when dealing with accurate spectral fitting of the X-ray
background before concluding.

%%%%%%%%%%%%%%%%%%%%%%%%%%%%%%%%%%%%%%%%%%%%%%%%%%%%%%%%%%%%%%%%%%%%%%%%%%%%%
\section{The model}
%%%%%%%%%%%%%%%%%%%%%%%%%%%%%%%%%%%%%%%%%%%%%%%%%%%%%%%%%%%%%%%%%%%%%%%%%%%%%
\label{mod}
We use the thermal Comptonization code of Haardt (1994, hereafter
H94). This code computes the angle--dependent spectrum of the
disk--corona system in a slab configuration using an iterative scattering
method, where the scattering anisotropy is taken into account only up to
the first scattering order. { It also computes, separatly, the
reflection components obtained including or not the comptonization
effects. They will be denoted $R_{\rm dir}$ and $R_{\rm comp}$,
respectively, in the following.} The continuum will be simply called
$C$.\\

\begin{figure*}
%\begin{figure*}[t]
\begin{tabular}{cc}
\psfig{width=0.45\textwidth,file=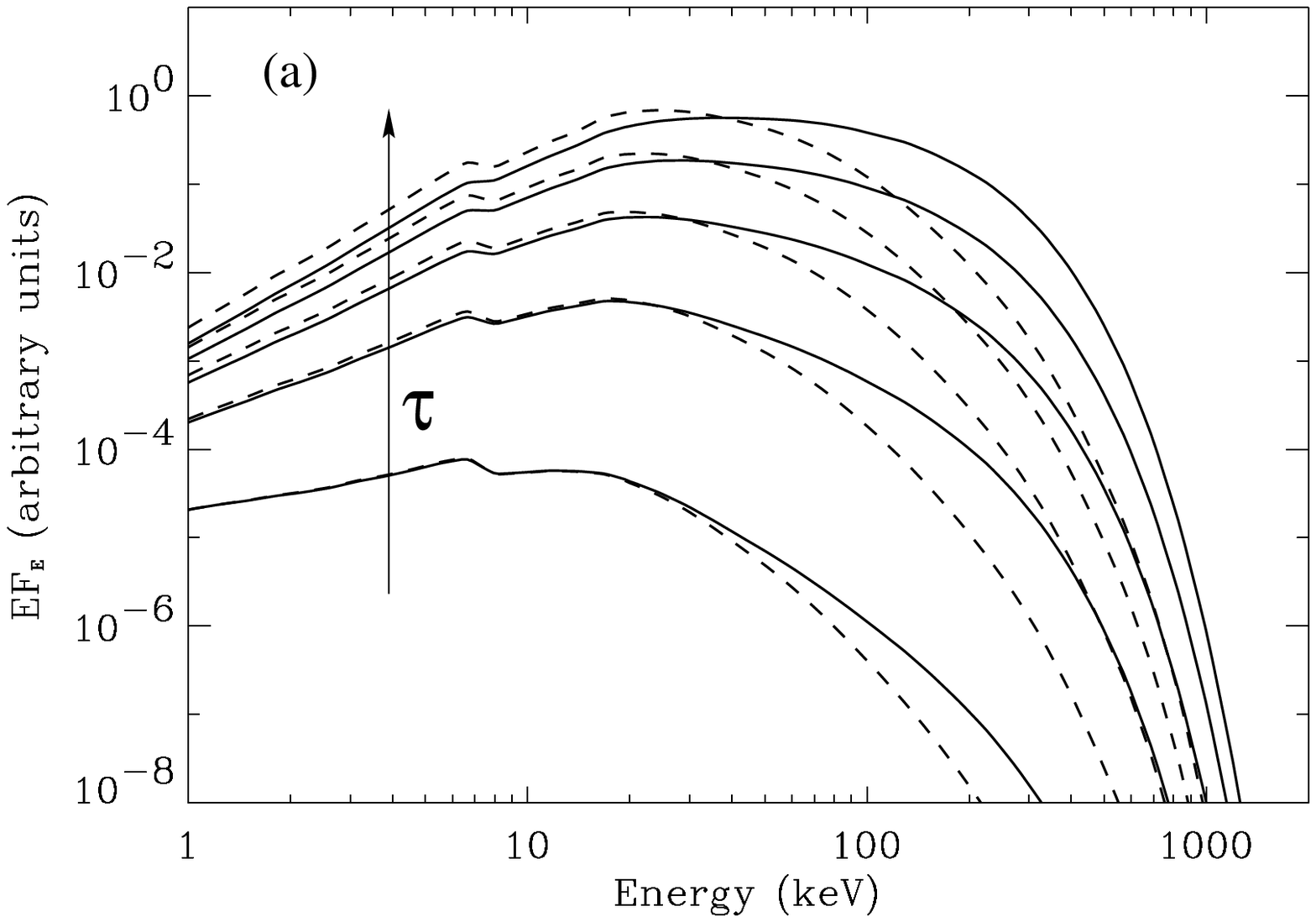}&
\psfig{width=0.45\textwidth,file=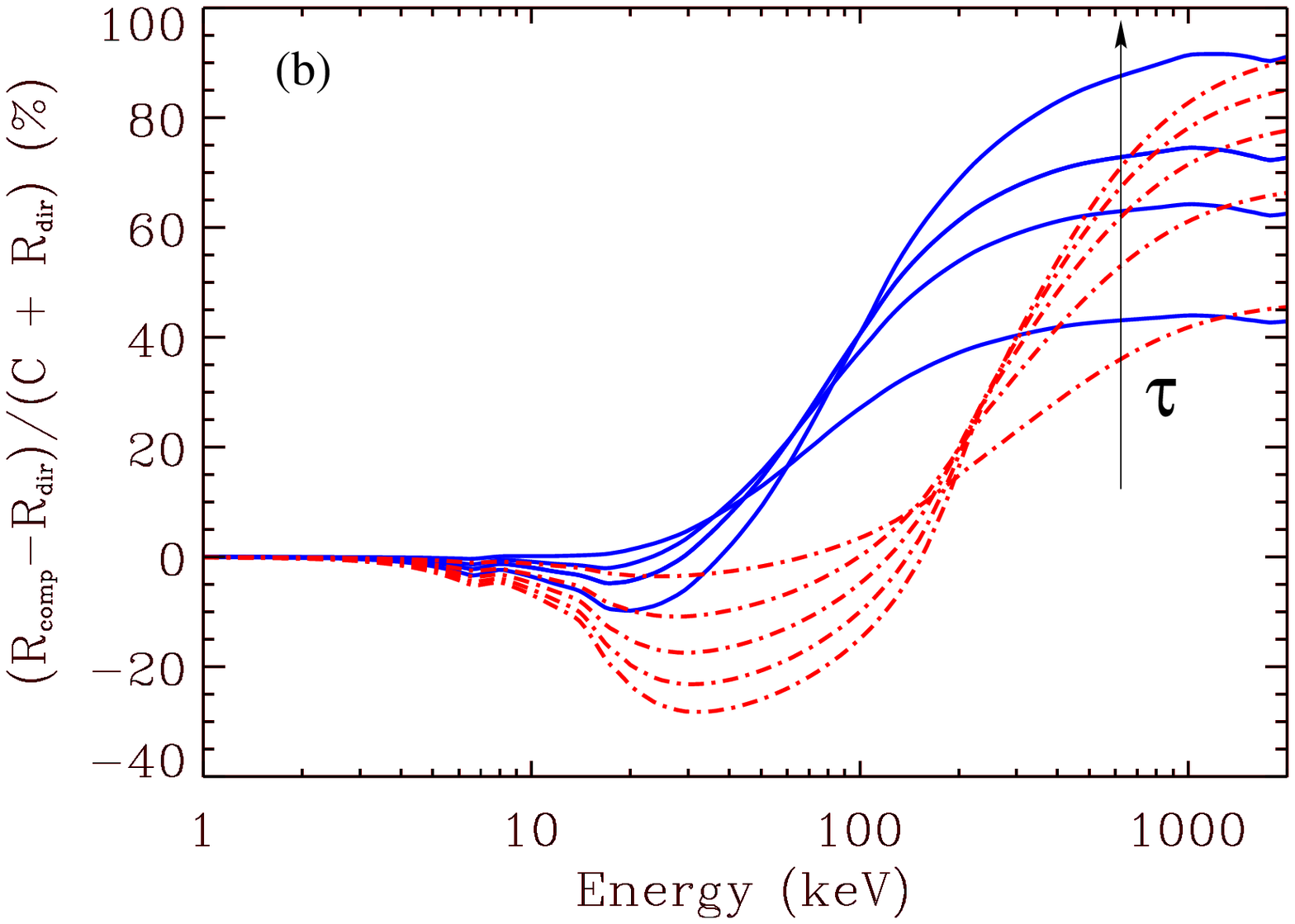}
\end{tabular}
\caption{(a) The comptonized (solid lines) and uncomptonized (dashed
lines) reflection humps, $R_{comp}$ and $R_{dir}$, for different values
of the coronal optical depth, the temperature being fixed to $kT_{\rm
e}$=50 keV. From bottom to top, $\tau$=0.1, 0.3, 0.5, 0.7 and 0.9. (b)
Deviations (in \%) between the outgoing spectra. The solid line and
dot-dashed lines correspond to $kT_e$ = 50 keV and 350 keV respectively.}
\label{fig1}
\end{figure*}
The reflection component $R_{\rm dir}$ is computed following White,
Lightman \& Zdziarski \shortcite {whi88} and Lightman \& White \shortcite
{lig88}; thus, the spectral shape of the reflected photons is averaged
over angles. It is well-known that the actual shape of the reflection
component does depend on the viewing angle, especially at high
energy. Such angular distortion is relatively complex and is a function
of the photon energy. The main effects occurs in hard X-rays and soft
$\gamma$-rays bands, where the reflected spectra strongly hardens with
increasing viewing angle \cite{hua92,mag95}. We have checked {\it a
posteriori} (cf. section \ref{varmu} and Fig. \ref{comppex}) that the
effects produced by such angular dependence are always smaller than the
one we discuss here.  Consequently, in the rest of the paper, we will
normalize the spectrum at different viewing angles by just multiplying
the reflection component by the angular function
\begin{eqnarray}
f(\mu)&=&\frac{3\mu}{4}\left[(3-2\mu^2+3\mu^4)
\ln\left(1+\frac{1}{\mu}\right)+\right.\nonumber\\
 & &\left.(3\mu^2-1)\left(\frac{1}{2}-\mu\right)\right],
\label{ryl}
\end{eqnarray}
neglecting any energy dependence (see Ghisellini, Haardt \& Matt 1994 for
details). It is worth noting that the shape of the reflection component
averaged over angles is very similar to the real shape expected with an
inclination angle of 60$^{\circ}$.\\

The effect of Comptonization on the reflection component will depend
mainly on three physical parameters: the temperature $kT_{\rm e}$ and the
vertical optical depth $\tau$ of the corona and the inclination angle $i$
(or its cosine $\mu=\cos i$). The last two parameters may be combined to
give an ``effective''optical depth $\tau_{\mu}=\tau/\mu$, which is the
line-of-sight optical depth that a photon, emitted at the surface of the disk,
has to cross in order to escape from the corona without being comptonized
and reach an observer at the viewing angle $i$.\\

The soft temperature of the cold matter $kT_{\rm bb}$, does not play an
important role in the problem we study here. In fact, in an anisotropic
geometry, as the one we are dealing with, the spectrum emitted by the
corona depends on such temperature because of the presence of an
anisotropy break \cite{ste95,sve96,haa97,pet00}. Such a break is due to
the apparent reduced contribution of the first Compton scattering order,
mainly emitted backward towards the disk, to the outgoing flux. The
spectrum is then better approximated by a convex broken power law where
the break energy depends mainly on $kT_{\rm e}$ and $kT_{\rm
bb}$. Changing $kT_{\rm bb}$ will thus modify the shape of the X-ray
spectrum impinging on the cold matter and consequently the shape of the
reflection components $R_{\rm dir}$ and $R_{\rm comp}$. However, we have
checked that these effects are negligible in comparison to the ones we
are looking at. We will thus fix $kT_{\rm bb}$ to a constant value of 10
eV in the following.\\

\begin{figure*}
%\begin{figure*}[t]
\begin{tabular}{cc}
\psfig{width=0.45\textwidth,file=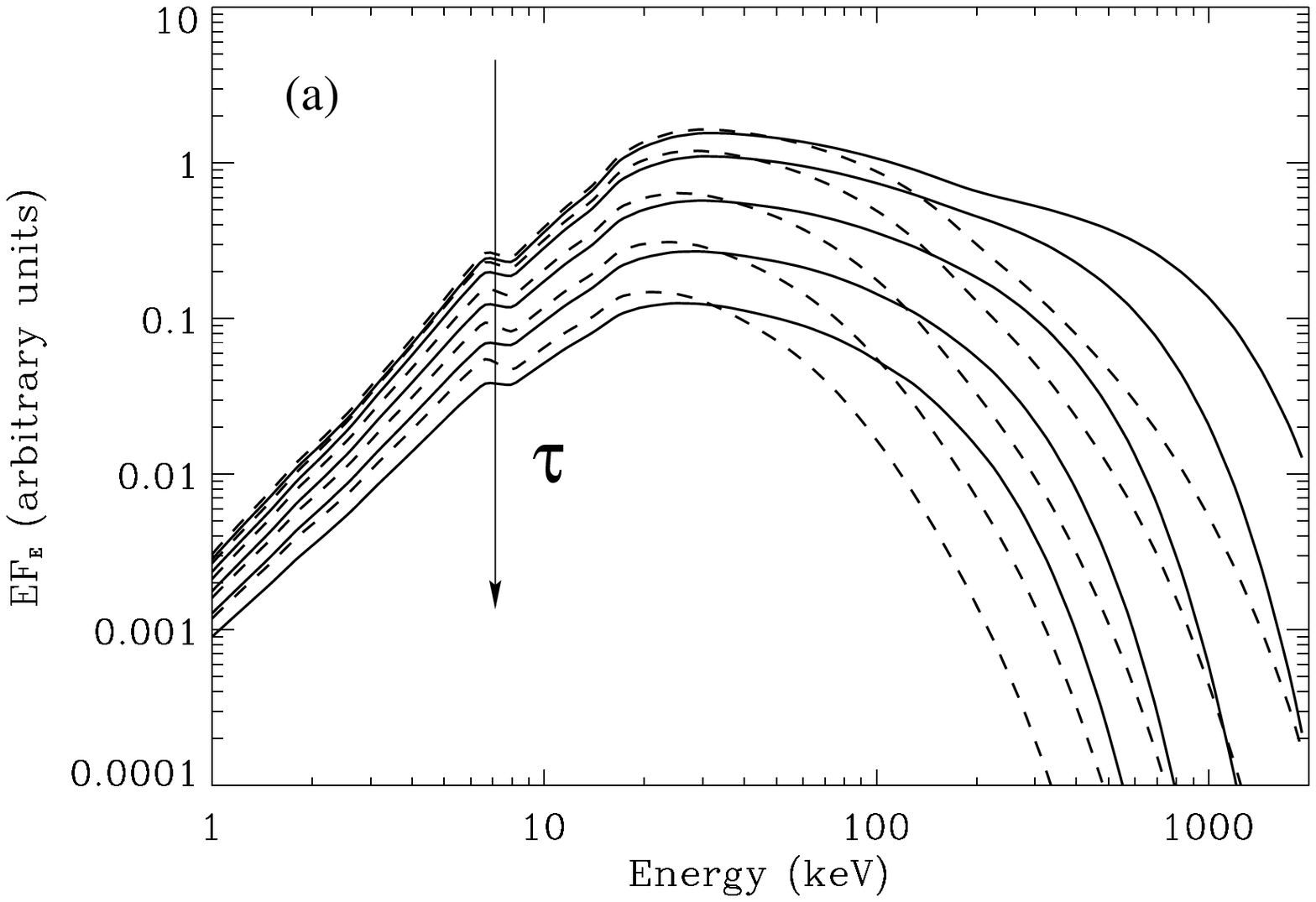}&
\psfig{width=0.45\textwidth,file=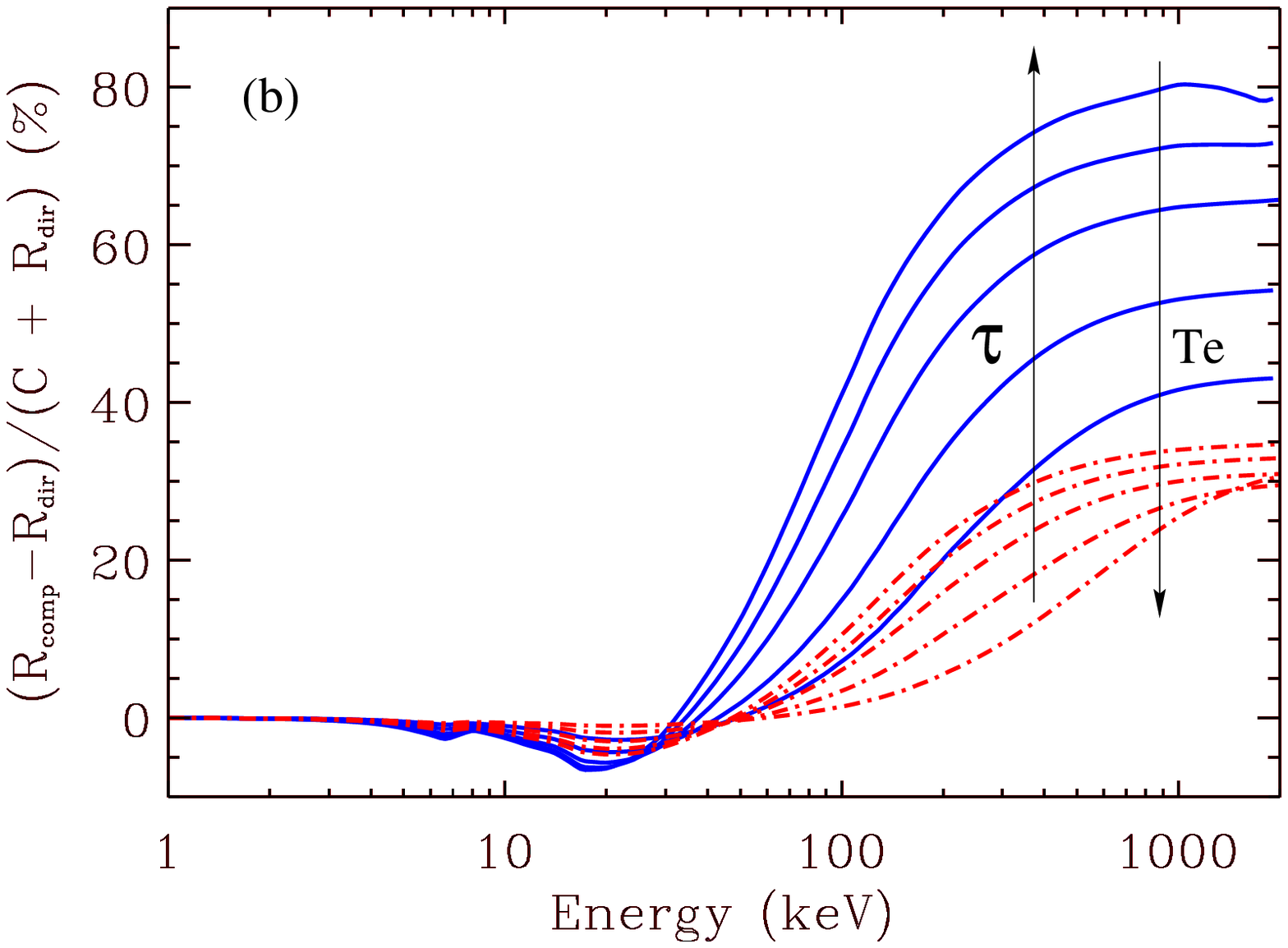}
\end{tabular}
\caption{(a) As in Fig. \ref{fig1}a, but for a constant Compton parameter
$y$ equal to 0.6, i.e. consistent with a slab corona in radiative
equilibrium above a passive accretion disk. From bottom to top
$\tau$=0.1, 0.2, 0.35, 0.5 and 0.64 and $kT_e$=240, 150, 90, 65 and 50
keV. (b) As in Fig. \ref{fig1}b but for a constant Compton parameter $y$
equal to 0.6 (solid line) and 2 (dot-dashed lines). The latter case is
consistent with a hemispherical corona in radiative equilibrium above a
passive accretion disk.}
\label{fig2}
\end{figure*}

%%%%%%%%%%%%%%%%%%%%%%%%%%%%%%%%%%%%%%%%%%%%%%%%%%%%%%%%%%%%%%%%%%%%%%%%%%%%
\section{Comptonization effects on the reflection hump shape}
%%%%%%%%%%%%%%%%%%%%%%%%%%%%%%%%%%%%%%%%%%%%%%%%%%%%%%%%%%%%%%%%%%%%%%%%%%%%
\label{compeff}
%%%%%%%%%%%%%%%%%%%%%%%%%%%%%%%%%%%%%%%%%%%%%%%%%%%%%%%%%%%%%%%%%%%%%%%%%%%%
\subsection{Dependence on the physical parameters of the corona}
%%%%%%%%%%%%%%%%%%%%%%%%%%%%%%%%%%%%%%%%%%%%%%%%%%%%%%%%%%%%%%%%%%%%%%%%%%%%
\label{varcor}
In this section we will firstly discuss the effect of the Comptonization
in the corona on the shape of the reflection component varying the
optical depth and/or the temperature of the corona, but keeping fixed the
viewing angle (here $\mu=0.9$). According to thermal Comptonization
theory, the temperature of the corona is directly related to the average
fractional energy change that a photon undergoes per scattering, whereas
$\tau$ gives an estimate of the mean number of scattering events. Both
parameters play thus a major role in the Comptonization process.

%%%%%%%%%%%%%%%%%%%%%%%%%%%%%%%%%%%%%%%%%%%%%%%%%%%%%%%%%%%%%%%%%%%%%%%%%%%%
\subsubsection{Varying $\tau$ at fixed coronal temperature}
%%%%%%%%%%%%%%%%%%%%%%%%%%%%%%%%%%%%%%%%%%%%%%%%%%%%%%%%%%%%%%%%%%%%%%%%%%%%
\label{tempfix}
We will first suppose here that the temperature of the corona is fixed,
equal to 50 keV, while $\tau$ varies. We have plotted in Fig. \ref{fig1}a
the reflection shapes of $R_{\rm dir}$ and $R_{\rm comp}$ { (in dashed
and solid line respectively)} for different values of $\tau$. \\

The increase of the optical depth hardens the X-ray primary spectrum,
thus modifying the intrinsic shape of the reflection component. It thus
explains the hardening of the uncomptonized reflection $R_{\rm dir}$
between $\tau$=0.1 and $\tau$=0.9. Increasing $\tau$ also magnifies the
effect of the Comptonization. The larger $\tau$, the larger the
probability of a photon to be comptonized. In this process the reflected
photons are shifted towards higher energies and an increasing deviation
between $R_{\rm dir}$ and $R_{\rm comp}$ is seen, at energies below and
above $\sim kT_{\rm e}=$50 keV, for increasing $\tau$.\\

The optical depth of the corona also controls the relative intensity of
the different Compton scattering orders. The larger $\tau$, the higher
the intensity of the different orders producing a hardening of the
comptonized shape at high energy (above $\sim$10--30 keV) as observed in
Fig. \ref{fig1}a.\\

To estimate quantitatively these effects on the total outgoing spectrum,
we have plotted in Fig. \ref{fig1}b (solid lines) the deviations (in \%)
between the total spectrum $(C+R_{\rm comp})$, expected if all the
reflected photons cross the comptonizing corona before being observed,
and the total one $(C+R_{\rm dir})$ predicted when the Comptonization of
the reflection hump is not taken into account. The most important effects
(variations of $>$ 50\%) occur at very high energies (above 100
keV). They are due to the hardening of $R_{\rm comp}$ for large optical
depths. For the larger optical depth case considered ($\tau=0.9$), a
factor of $\sim$2 is expected near 1 MeV. At these energies, $R_{\rm
comp}$ tends to have the same shape as the primary continuum and the
fractional deviations between the two spectra attain a maximum.\\

The Comptonization produces also smaller ($<$ 10\%) differences between
$R_{\rm dir}$ and $R_{\rm comp}$ near 10 keV.  The larger one is still
produced in the high optical depth case, as expected. It is worth noting
that the X--ray detectors are generally well sensitive in the 10--20 keV
energy range, which can help constraining the degree of Comptonization.\\

Also plotted in Fig. \ref{fig1}b are the results obtained for a coronal
temperature of 350 keV (dot-dashed lines). We can still observe
differences between $R_{\rm dir}$ and $R_{\rm comp}$ below and above
$\sim kT_e$.  However, such deviations are larger than in the previous
case, at least at low energy (below 100 keV), since the average
fractional energy change of a scattering photon increases with $kT_{\rm
e}$. These deviations reach $\sim$40\% near 20-30 keV for
$\tau=0.9$. Above 100 keV, they roughly behave as in the small
temperature case.

\begin{figure*}
%\begin{figure*}[t]
\begin{tabular}{cc}
\psfig{width=0.45\textwidth,file=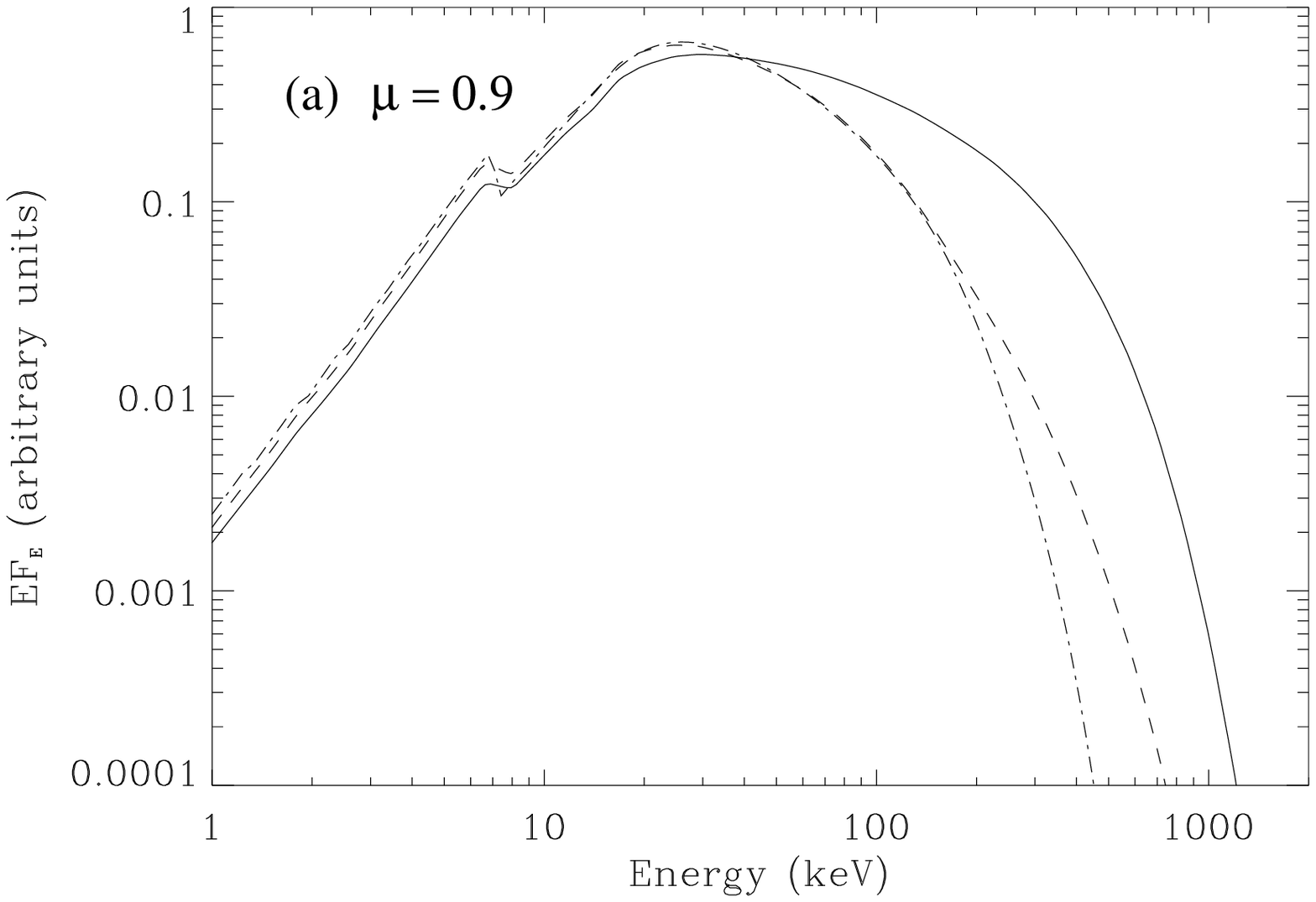}&
\psfig{width=0.45\textwidth,file=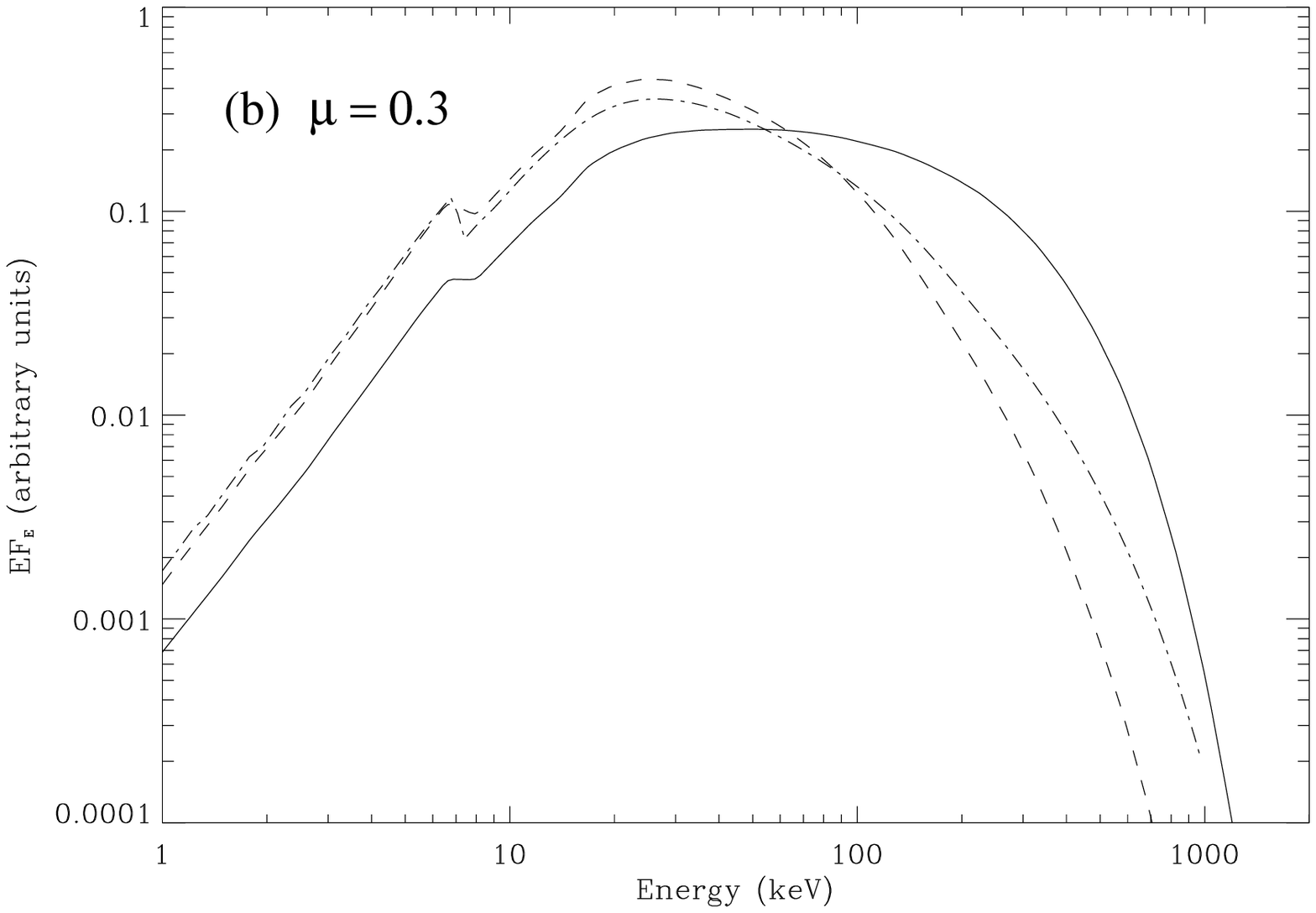}
\end{tabular}
\caption{Plots of $R_{comp}$ (solid line), $R_{dir}$ (dashed line) and
$R_{pexrav}$ (dot-dashed line) at two different viewing angles: (a)
$\mu$=0.9 and (b) $\mu$=0.3.}
\label{comppex}
\end{figure*}
\begin{figure*}
%\begin{figure*}[t]
\begin{tabular}{cc}
\psfig{width=0.45\textwidth,file=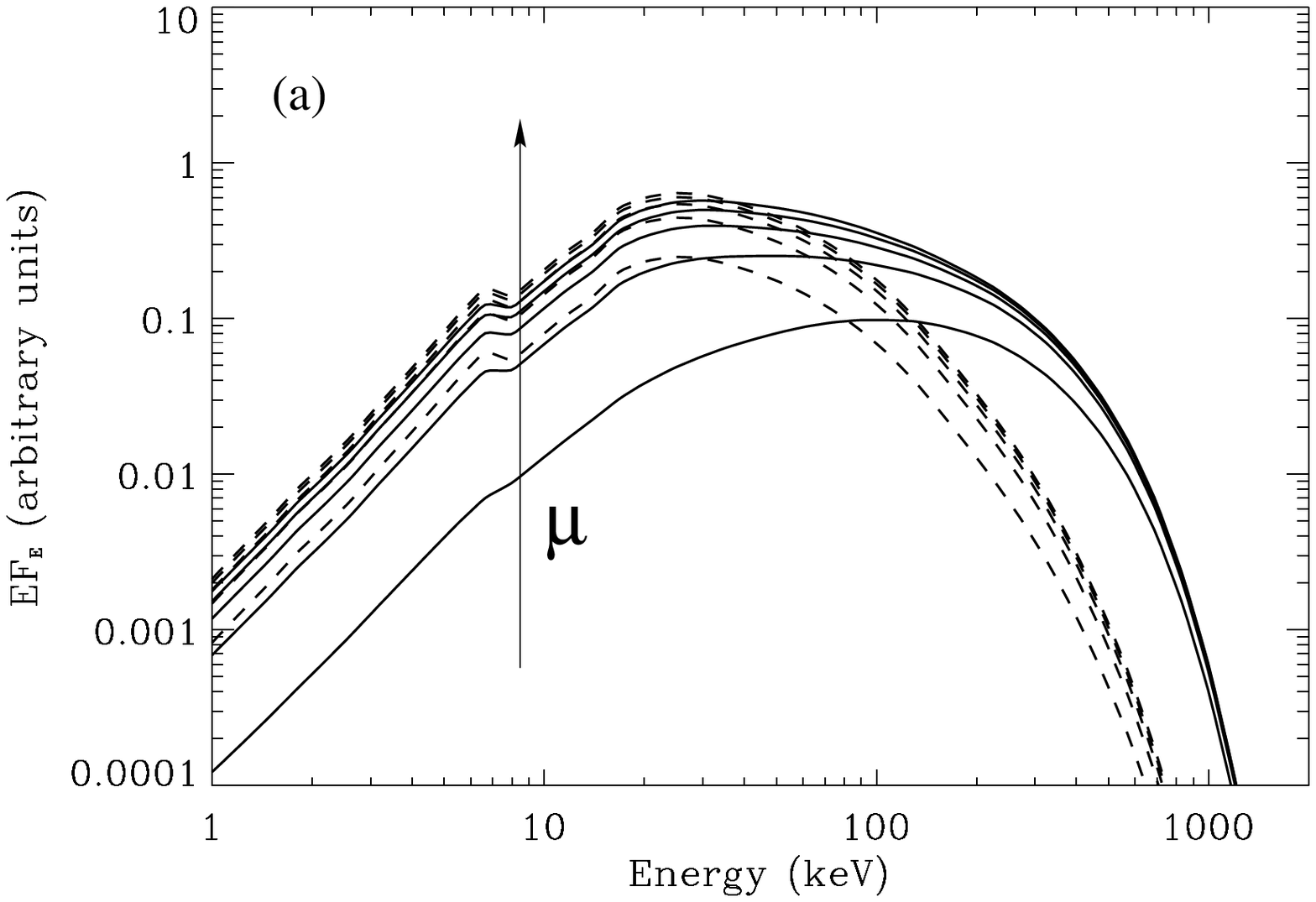}&
\psfig{width=0.45\textwidth,file=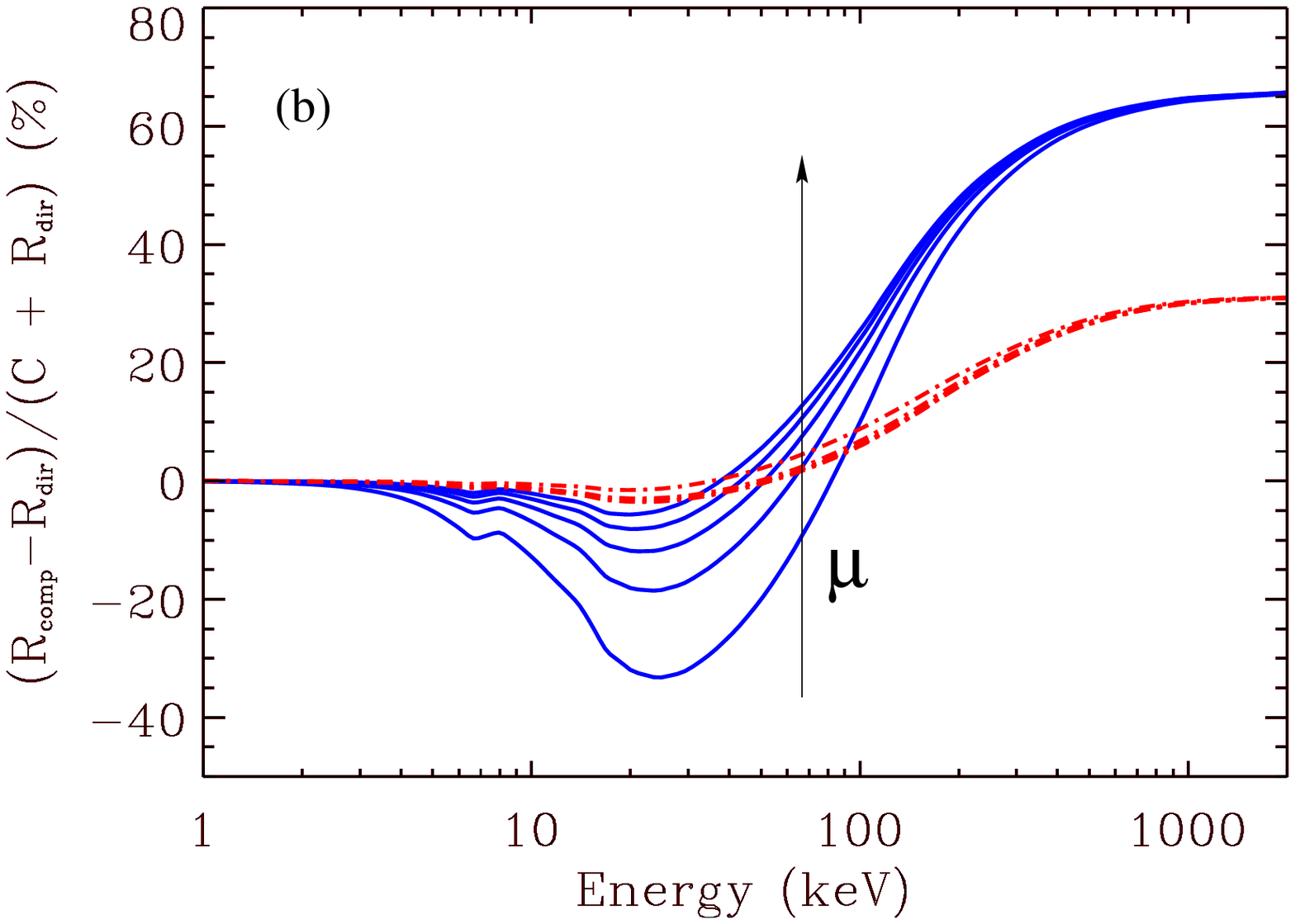}
\end{tabular}
\caption{(a) The comptonized (solid lines) and uncomptonized (dashed
lines) reflection humps, $R_{comp}$ and $R_{dir}$, for different viewing
angle.  From bottom to top $\mu$=0.1, 0.3, 0.5, 0.7 and 0.9. The optical
depth and temperature of the corona are fixed to 0.35 and 90 keV
respectively. (b) Deviations (in \%) between the outgoing spectra. In
dot-dashes line the case of an hemispherical corona with $\tau$=0.35 and
$kT_{\rm e}$=200 keV.}
\label{fig4}
\end{figure*}
%%%%%%%%%%%%%%%%%%%%%%%%%%%%%%%%%%%%%%%%%%%%%%%%%%%%%%%%%%%%%%%%%%%%%%%%%%%%
\subsubsection{Varying $\tau$ and $T_{\rm e}$ at fixed Compton parameter}
%%%%%%%%%%%%%%%%%%%%%%%%%%%%%%%%%%%%%%%%%%%%%%%%%%%%%%%%%%%%%%%%%%%%%%%%%%%%
\label{fixy} In the case of a disk--corona system, such as the one we 
are dealing with, we expect the comptonizing region and the source of
soft photons to be {\it coupled}, as the optically thick disk necessarily
reprocesses and re-emits part of the comptonized flux as soft photons
which are the seeds for Comptonization. In this case, the system can only
be in equilibrium if the temperature and optical depth satisfy a precise
relation \cite{haa91}. That in turn corresponds to roughly constant
Compton $y$ parameters, defined as:
\begin{equation}
y=4\left(\frac{kT_{\rm e}}{m_{\rm e}c^2}\right)\,
\left[1+4\left(\frac{kT_{\rm e}}{m_{\rm e}c^2}\right)\right] \tau (1+\tau),
\end{equation}
which is about $0.6$ in a slab geometry when all the accretion power is
released in the corona \cite{haa93a,ste95,sve96}.

In the case of a constant Compton parameter, changes in $\tau$ will be
necessarily accompanied by changes in $kT_{\rm e}$, namely a larger
optical depth will require a smaller coronal temperature.  This results
in three effects: first, the photon index of the primary continuum
correlates with $\tau$ \cite{haa97}, whereas these two parameters are
anti-correlated if the temperature is kept fixed (cf. section
\ref{tempfix}). Second, the variation of the coronal temperature modifies
the high energy cut--off of the primary continuum, which scales roughly
as $2kT_{\rm e}$.  Consequently, it slightly modifies the high energy
part of the reflection component $R_{\rm dir}$. Finally, the change of
$kT_{\rm e}$ modifies the mean photon energy gain per scattering.\\

These different effects produce a hardening of $R_{\rm comp}$ at high
energy when $\tau$ decreases as can be seen in Fig. \ref{fig2}a. This
behavior is thus the opposite of what we obtained when the temperature is
fixed. However, as shown in Fig. \ref{fig2}b (solid lines), the
deviations between the outgoing spectra closely resemble those shown in
Fig. \ref{fig1}b, becoming larger for larger optical depth (i.e. smaller
temperature) and reaching a plateau at high energies (above 200 keV). For
comparison, we have also plotted the deviation curves obtained in the
case of a hemispherical geometry in radiative equilibrium (dot-dashed
lines). In this case, the Compton parameter is of the order of 2 (if all
the accretion power is released in the corona, see Stern et al. 1995). We
see that the deviations are smaller (by about a factor of 2) than in the
slab case. This is simply due to the fact that in hemispherical geometry,
the corona does not cover all the cold disk. Consequently, only a part of
the reflected photons have to cross the comptonizing plasma before being
observed, reducing the effects of Comptonization.

%%%%%%%%%%%%%%%%%%%%%%%%%%%%%%%%%%%%%%%%%%%%%%%%%%%%%%%%%%%%%%%%%%%%%%%%%%%%
\subsection{Dependence on the inclination angle}
%%%%%%%%%%%%%%%%%%%%%%%%%%%%%%%%%%%%%%%%%%%%%%%%%%%%%%%%%%%%%%%%%%%%%%%%%%%%
\label{varmu}
In this section we fix the temperature and optical depth of the corona to
90 keV and 0.35 respectively. Those values correspond to a Compton
parameter $y$ close to 0.6, i.e. consistent with a slab corona in
radiative equilibrium above a passive accretion disk (cf. previous
section). We focus here on the dependence of the Comptonization effects
on the system viewing angle.\\

As already mentioned in the introduction, the actual shape of the
reflection hump depends on $\mu$ in a relatively complex manner which is
a function of the photon energy. We recall that this angular dependence
is neglected in our computation of $R_{\rm dir}$, whose shape is constant
and corresponds to the angular averaged one. We have plotted in
Figs. \ref{comppex}a and \ref{comppex}b the reflection component $R_{\rm
dir}$, computed by our code (dashed line), and the one computed following
Magdziarz \& Zdziarski \shortcite {mag95}, obtained with the PEXRAV model
of XSPEC, and noted $R_{\rm pexrav}$ (dot-dashed line), for two different
inclination angles $\mu=0.9$ (Fig. \ref{comppex}a) and $\mu=0.3$
(Fig. \ref{comppex}b). $R_{\rm pexrav}$ is computed using angle dependent
Green functions and takes properly into account the angular distortion of
the Compton reflection. The continuum assumed to compute $R_{\rm pexrav}$
corresponds to the best cut-off power law fit of the spectrum emitted by
our slab corona model, that is a photon index $\Gamma\simeq 2.2$ and a
high energy cut--off $E_c\simeq$ 250 keV. The primary spectra impinging
on the disk are thus roughly similar in the two models. We see in
Fig. \ref{comppex}a and \ref{comppex}b that the main differences between
$R_{\rm dir}$ and $R_{\rm pexrav}$ appears above 10 keV. $R_{\rm pexrav}$
also hardens for increasing $i$ (i.e. decreasing $\mu$).  For comparison,
we have also plotted on these figures the comptonized reflection $R_{\rm
comp}$ computed with our model. It is evident that the effects of the
Comptonization on the reflection hump are generally more important than
the intrinsic angular ones, especially at high energies.  Therefore, in
the following we will simply use Eq. (\ref{ryl}) to rescale the amplitude
of the reflection component as a function of $\mu$.\\
 
Keeping in mind the limitations of the above approximation, we have
plotted in Fig. \ref{fig4}a, the different shapes of $R_{\rm comp}$ and
$R_{\rm dir}$ for different values of the viewing angles.  Because an
increase of $i$ will correspond to an increase of the effective optical
depth $\tau_{\mu}$, we expect to observe, for increasing $i$, the same
effects we observed for increasing $\tau$ (cf. section
\ref{varcor}). However, the main difference is that now the shape of the
primary spectrum impinging on the accretion disk is constant.\\

The differences between $R_{\rm comp}$ and $R_{\rm dir}$ are relatively
strong (cf. Fig.  \ref{fig4}a).  In our case, $\tau_{\mu}$ can easily
reach values of the order of 2 or 3 for large inclination. It thus
results in large mean photon energy shifts and hardening of $R_{\rm
comp}$ at high energy. The deviations between $R_{\rm dir}$ and $R_{\rm
comp}$ are reported in Fig. \ref{fig4}b. They may be as large as
$\sim$30\% in the 10--50 keV range for $\mu$=0.1 whereas they are of the
order of a few for nearly face-on configuration.\\

Above 100 keV the different shapes overlap near $2kT_e$ the larger energy
that a photon may reach by Comptonization in the corona. It is worth
noting that this overlapping is a direct consequence of our treatment of
the angular dependence of the reflection hump. Actually, the reflection
shapes for different inclination angles would be different especially at
high energies. The deviations would however not strongly depend on the
real reflection shape and we see that they are of the order of 60\% near
300 keV for a slab geometry. They reach only 30\% for an hemispherical
corona since the Comptonization effects are less important in this case
(cf. section \ref{fixy}).\\

{ We note however that the Comptonization effects beyond $\sim$100 keV
may not be easily detected. Firstly because the high energy instruments
are generally not very sensitive beyond $\sim$100 keV and thus prevent
any good constrains on the spectral shape. Secondly, any deviations from
the ideal case studied here like, for instance, a stratified temperature
corona, may distort the derived reflected shape, especially at high
energy, and complicate the spectral analysis. The effects at lower energy
(near $\sim$20 keV where the instrumental sensitivity is very good) would
certainly be more relevant for any estimation of the comptonization
effects.  }

%%%%%%%%%%%%%%%%%%%%%%%%%%%%%%%%%%%%%%%%%%%%%%%%%%%%%%%%%%%%%%%%%%%%%%%%%%%%
\section{Comptonization effects on the iron line}
%%%%%%%%%%%%%%%%%%%%%%%%%%%%%%%%%%%%%%%%%%%%%%%%%%%%%%%%%%%%%%%%%%%%%%%%%%%%
\label{eweffect} The equivalent width (EW) of the fluorescent Fe K$\alpha$
line produced by cold matter surrounding the hot corona depends on
different parameters, as the elemental abundance of the reflecting
matter, its inclination angle as seen by the observer, and the ionization
state of the reflecting surface layers (see Fabian et al. 2000 for a
review). It also depends on the geometry of the corona + cold matter
configuration (i.e. on the solid angle subtended by the cold matter as
seen by the X--ray source). The Comptonization of the iron line is
another process which may affect the measurement of this EW as pointed by
Haardt et al. \shortcite {haa93b} and discuss by Matt et al. \shortcite
{mat97}.  We try to quantify a bit more this effect here.\\

Assuming a slab corona geometry above an accretion disk of neutral matter
and solar abundance, the dependence of the EW on $\mu$ can be
approximated by the simple formula:
\begin{eqnarray}
EW_{\rm comp}(\mu)&=&\frac{EW_{\rm dir}(\mu=1)}{\ln{2}}\mu
        \ln\left(1+\frac{1}{\mu}\right)e^{-\tau_{\mu}}\label{EWeq}\\
        &=&EW_{\rm dir}(\mu)e^{-\tau_{\mu}}. \nonumber
\end{eqnarray}
The term before the exponential is an approximated formula, given by
Ghisellini et al. \shortcite {ghi94}, of the angular dependence of the
equivalent width due to limb darkening, $EW_{\rm dir}(\mu=1)$
corresponding to the face-on EW of the uncomptonized line. The
exponential term $e^{-\tau_{\mu}}$ gives the probability of a line photon
to cross the corona and reach an observer at the viewing angle $i$
without being comptonized. The comptonized ones will be scattered around
in the underlying continuum and will not contribute to the line flux
anymore.\\

\begin{figure}
%\begin{figure}[h]
\psfig{width=\columnwidth,file=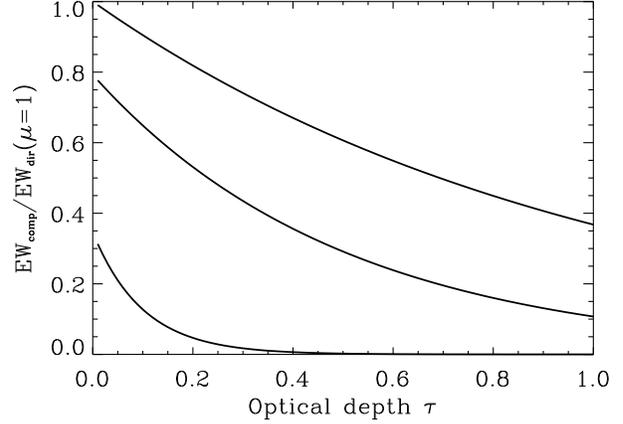}
\caption{Reduced equivalent width $EW_{\rm comp}(\mu)/EW_{\rm
dir}(\mu=1)$ of the comptonized iron line in function of the optical
depth of the corona for different inclination angles. From top to bottom
$\mu$ = 1, 0.5 and 0.1}
\label{fig5}
\end{figure}
We have plotted in Fig. \ref{fig5} the reduced equivalent width $EW_{\rm
comp}(\mu)/EW_{\rm dir}(\mu=1)$ as a function of the optical depth of the
corona and for different viewing angles.  We see that the effects of
Comptonization are very strong, especially at large inclination
angles. For $\mu=0.9$ the EW is reduced by a factor $\sim$3 when $\tau$
varies between 0 and 1 whereas it becomes rapidly negligible for
$\tau>0.5$ if $\mu=0.1$.{ We thus expect these modifications of the
Iron line EW to be a lot more observable than the spectral changes of the
reflection hump detailed in section \ref{compeff}}\\

%%%%%%%%%%%%%%%%%%%%%%%%%%%%%%%%%%%%%%%%%%%%%%%%%%%%%%%%%%%%%%%%%%%%%%%%%%%%
\section{Application to fitting procedures}
%%%%%%%%%%%%%%%%%%%%%%%%%%%%%%%%%%%%%%%%%%%%%%%%%%%%%%%%%%%%%%%%%%%%%%%%%%%%
%%%%%%%%%%%%%%%%%%%%%%%%%%%%%%%%%%%%%%%%%%%%%%%%%%%%%%%%%%%%%%%%%%%%%%%%%%%%
%\subsection{Simulations}
%%%%%%%%%%%%%%%%%%%%%%%%%%%%%%%%%%%%%%%%%%%%%%%%%%%%%%%%%%%%%%%%%%%%%%%%%%%%
\label{simu} In section \ref{compeff} we have discussed, in an admittedly
qualitative way, the changes in the observable reflection hump due to the
presence of a comptonizing corona above the cold reflector (the accretion
disc). Here we will try to give a more quantitative estimate of such an
effect. To do so, we have to rely on some simple and straightforward
measure of the reflection component, and compare such a measure in the
two cases of a bare disc and of a disc engulfed in a hot optically thin
corona.\\

The most widely used model to fit the hard X-ray continuum of Seyfert
galaxies and GBHC comprises a cut--off power law and an uncomptonized
reflection (the so-called PEXRAV model of XSPEC, Magdziarz \& Zdziarski
1995). { This model is known to be a poor approximation to the true
Compton continuum since it does not reproduce the (possible) anisotropy
break (cf. section 2) and may give (mainly for high, i.e. larger than 1,
optical depth) a poor modelisation of the high energy cut--off
shape. However, it can help to quantify the effects of a comptonizing
corona on the appearance of the hard X-ray continuum. For this purpose we
have proceeded in the following way.} We have first created two sets of
simulated spectra using the thermal Comptonization code of H94 in a slab
geometry.  In the first set we have properly taken into account the
Comptonization of the secondary components in the corona, while in the
second one we have neglected such effect. Then we have fitted our
simulated spectra between 1 and 500 keV with the PEXRAV model.  We have
assumed different viewing angles and different values for coronal
temperature and optical depth, chosen as to keep $y=0.6$. For every
simulated spectrum, our fitting procedure gives the { best fit} value
of the spectral index, the high energy cut-off and of the reflection
fraction. We will denote $R_1$ and $E_{c,1}$, respectively, the value of
the reflection fraction and of the high energy cut-off obtained for the
comptonized model, and $R_2$ and $E_{c,2}$ the ones obtained for the
uncomptonized case. { The ratio $R_1/R_2$ and $E_{c,2}/E_{c,1}$ can
then help to quantitatively estimate the modifications of the best fit
parameter values due to the comptonization effects}. \\

\begin{figure}
%\begin{figure}[h]
\psfig{angle=0,width=\columnwidth,file=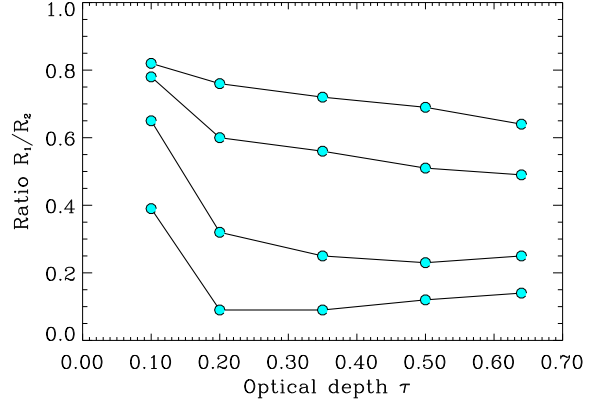}
\caption{Ratio $R_1$/$R_2$ versus the coronal optical depth for different
viewing angles but keeping the Compton parameter $y$ equal to 0.6. From
top to bottom $\mu$=0.9, 0.5, 0.2 and 0.1}
% ($\tau$,$kT_e$). From top to bottom we have
%($\tau$, $kT_e (keV)$)=(0.1, 480);(0.2, 296);(0.35, 187);(0.5,
%133);(0.64; 102);(0.8, 81).}
\label{r1r2}
\end{figure}

In Fig. \ref{r1r2} we show the ratio $R_1/R_2$ as a function of the
optical depth and for different viewing angles $i$.  As expected, due to
the smoothing effect of the comptonizing corona on the reflection shape,
$R_1$ is always smaller than $R_2$, because the fitting procedure needs
smaller reflection normalization to fit the comptonized component.
Furthermore, their ratio decreases with increasing $i$ and/or optical
depth in the corona. This is in agreement with the results of section
\ref{compeff}.\\

%For small optical depth and small viewing angles (i.e. $\mu>0.8$), the
%effect of the anisotropy break in the spectrum on the fitted
%parameters is non negligible: when trying to fit the curved spectrum with
%a cut-off power-law, we obtain a lower value spectral index and,
%consequently, a lower cut-off energy. The combination of these two
%effects suppresses the amount of reflection and gives a small value for
%both $R_1$ and $R_2$ and reduce the measured ratio $R_1/R_2$.\\
 
\begin{figure}
%\begin{figure}[h]
\psfig{angle=0,width=\columnwidth,file=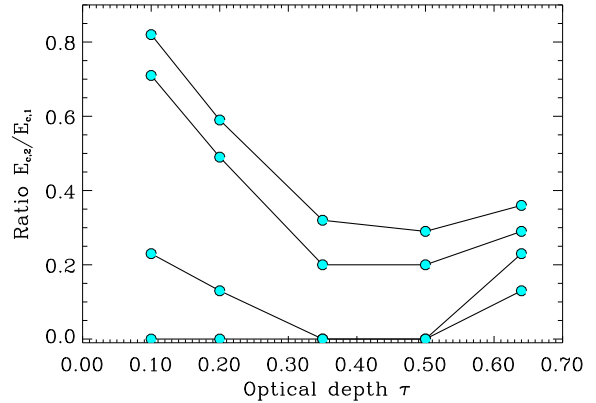}
\caption{Ratio $E_{c,2}$/$E_{c,1}$ versus the coronal optical depth, 
for different values of $\mu$. From top to bottom $\mu$=0.9, 0.5, 0.2 and
0.1.}   
\label{e1e2}
\end{figure}

The shifting of the reflection hump $R_{\rm comp}$ towards high energies
(as discussed in section 3) as a consequence of the Comptonization,
modifies also the high energy part of the spectra and the estimate of the
high energy cut--off so that it is always larger for the comptonized
case. The amount of such shift depends on the temperature and optical
depth of the corona (cf. Fig. \ref{fig2}a) and the higher the
temperature, the higher the energy of the scattered photons. At the same
time, however, in our simulations, higher coronal temperature
corresponding to lower optical depth, the effect of the anisotropy break
has to be taken into account. When trying to fit the curved spectrum with
a cut-off power-law, we obtain a lower value spectral index and,
consequently, a lower cut-off energy. The net result is shown in
Fig. \ref{e1e2}, where we have plotted the ratio $E_{c,2}/E_{c,1}$ of the
measured cut-off energies as functions of the optical depth for different
viewing angles. The ratio has a minimum for $\tau\simeq$0.4. At the
highest inclination angle ($\mu$=0.1--0.2) we are unable to constrain the
high energy cut-off in the comptonized case $E_{c,1}$, below 500 keV, so
we report for the ratio $E_{c,2}/E_{c,1}$=0.

{ The main qualitative conclusion of this section is that fitting
Comptonization models that take properly into account the comptonized
reflection by the usual cut--off power law + uncomptonized reflection
leads to an underestimation of the reflection normalization and also to
an overestimation of the high energy cut--off. Quantitatively, these
effects will of course strongly depend on the quality of the data,
especially at high energy (above 100 keV).}

\section{Observational consequences}
%%%%%%%%%%%%%%%%%%%%%%%%%%%%%%%%%%%%%%%%%%%%%%%%%%%%%%%%%%%%%%%%%%%%%%%%%%%%
\label{obscons}
%%%%%%%%%%%%%%%%%%%%%%%%%%%%%%%%%%%%%%%%%%%%%%%%%%%%%%%%%%%%%%%%%%%%%%%%%%%%
\subsection{The case of NGC 4258}
%%%%%%%%%%%%%%%%%%%%%%%%%%%%%%%%%%%%%%%%%%%%%%%%%%%%%%%%%%%%%%%%%%%%%%%%%%%%
%The nearby bright galaxy NGC 4258 is well known for the detection, from
%high resolution observations, of a water maser expected to be produce by
%the accreting mater spiraling down to the central engine (Claussen et
%al. \shortcite {cla84}; Watson et Wallin \shortcite {wat94}; Miyoshi et
%al. \shortcite {miy95}). This water maser enables a precise determination of
%the inclination of the disk and of the central binding mass i.e.:
%i=82$\pm$1 degrees (Heernstein et al. 1999) and M$\simeq 4\times 10^7
%M_{\sun}$ (Miyoshi et al. \shortcite {miy95}). This mass estimation in a
%relatively small region ($\sim$ 0.13 pc, Miyoshi et al. \shortcite {miy95}),
%strongly supports the presence of a supermassif black hole in this
%galaxy. Moreover, it is spectroscopically classified as a 1.9 Seyfert
%galaxy (Ho, Filippenko \& Sargent \shortcite {ho97}). The relatively broad and
%strongly polarized emission lines observed in this object (Wilkes et
%al. \shortcite {wil95}) also support the existence of a active nucleus in the
%central region of this galaxy.\\

>From the X--ray point of view, the nearby bright galaxy NGC 4258
possesses the general characteristic of the Seyfert class, that is: a
power law primary continuum with a photon index $\Gamma\simeq$ 2, a
relatively large amplitude variability in the 3-10 keV band on time
scales of a few tens of thousands of seconds and smaller ones
($\sim$20\%) on time scales of the order of an hour (Reynolds et
al. 2000, hereafter R00; Fiore et al. 2001, hereafter F01) and the
presence of a (narrow) iron line near 6.4 keV with an equivalent width
measured by ASCA of 107$^{+42}_{-37}$ eV (R00). The BeppoSAX observation
gives a more poorly constrained value for the EW of 85$\pm$65
eV. Furthermore, following R00, a broad iron line with an EW $<$ 200--300
eV may still be consistent with the ASCA data. On the other hand, the bad
signal to noise above 10--20 keV has prevented any good detection of a
possible reflection component.\\

NGC 4258 is well known for the detection, from high resolution
observations, of a water maser. It is expected to be produced by the
accreting gas spiraling down to the central engine
\cite{cla84,wat94,miy95}. The combination of the central mass estimate
(deduced from the water maser properties, Heernstein et al. 1999; Miyoshi
et al. 1995), and the NIR-to-X--ray luminosity suggest that NGC 4258 is
an AGN in a low state with an Eddington accretion rate of 0.0002
(F01). The presence of an ADAF seems however to be ruled out by the
recent BeppoSAX observations, on the base of both the measured X-ray
spectral shape and X-ray variability, and the most natural explanation of
the X-ray luminosity of this source may be in terms of Comptonization of
soft photons in a hot corona (F01). The inclination of the disk is large
and very well constrained i=82$\pm$1 degrees (Herrnstein et al. 1999). We
thus expect the Comptonization effects on the reflection components
coming from the disk to be relatively important, especially in the case
of a slab corona.\\

We can thus wonder, from the results obtained in the previous sections,
what constraints we can put on the origin of the reflection features
detected in this object. To this purpose, we have first try to estimate
the physical parameters (mainly the temperature and optical depth) of the
the corona. We have thus fit the BeppoSAX data (downloaded from the
BeppoSAX archive) with H94, assuming an inclination angle of $=$
82$^{\circ}$ (i.e. $\mu=0.14$). Since the intensity of the hard component
is strongly reduced by photoelectric absorption below 2--3 keV, we only
fit the data above 2.5 keV. Due to the low signal--to--noise ratio of the
spectrum, especially at high energy, and the lack of data to constrain
the soft photon temperature $kT_{\rm bb}$, $\tau$ and $kT_{\rm e}$ are
badly constrained, $\tau$ being smaller than 1 at the 90\% confidence
level for 2 parameters, with a best fit value of 0.05.\\

These large uncertainties still enable us to make some interesting
comments. First, let us assume, as considered by R00, that the narrow
iron line observed in NGC 4258 is from the accretion disk. Then it has to
originate at relatively large radii ($R_e >$ 100 Schwarzschild radii) in
order to produce the observed small line width. In this case, and since
the inclination angle of the disk is large (we assume that the
inclination of the inner part of the disk is also of 82$^{\circ}$,
i.e. the disk is not strongly warped) we expect part of the iron line
photons to be comptonized in the corona. Thus, as shown in section 4, the
EW may be significantly reduced in comparison to the one we expect
without a comptonizing corona. Let us suppose that all iron line photons
have to cross the corona before being observed (this would be the case if
the coronal dimension D is large i.e. $D>R_e$ as underlined by R00). The
observed EW of 100 eV would then correspond to an actual uncomptonized
face--on line EW (the EW$_{\rm dir}(\mu=1)$ parameter in
Eq. (\ref{EWeq})) larger than 300 eV ($\sim$500 eV for $\tau$=0.05 and
$\sim$40 keV for $\tau=0.5$ !). Recent estimates of the average narrow
iron line equivalent width observed in Seyfert 1 galaxies rather suggest
a value of the order $\sim$ 100 eV \cite{mat00,lub00}. This would
corresponds to an EW$_{\rm dir}(\mu=1)$ of $\sim$200 eV for $\tau$=0.5
and a viewing angle of 30$^{\circ}$ (as estimated by Nandra et al. 1997
for Seyfert 1). These (admittedly qualitative) computations rather
suggest that the narrow iron line in NGC 4258 does not suffer
Comptonization. The most simple explanation, with the assumption that the
X--ray emitting region has a disk--corona configuration, is that the
narrow iron line preferentially originates from matter not associated
with the accretion disk.\\

\begin{figure*}
%\begin{figure*}[t]
\begin{tabular}{cc}
\psfig{angle=0,width=0.45\textwidth,file=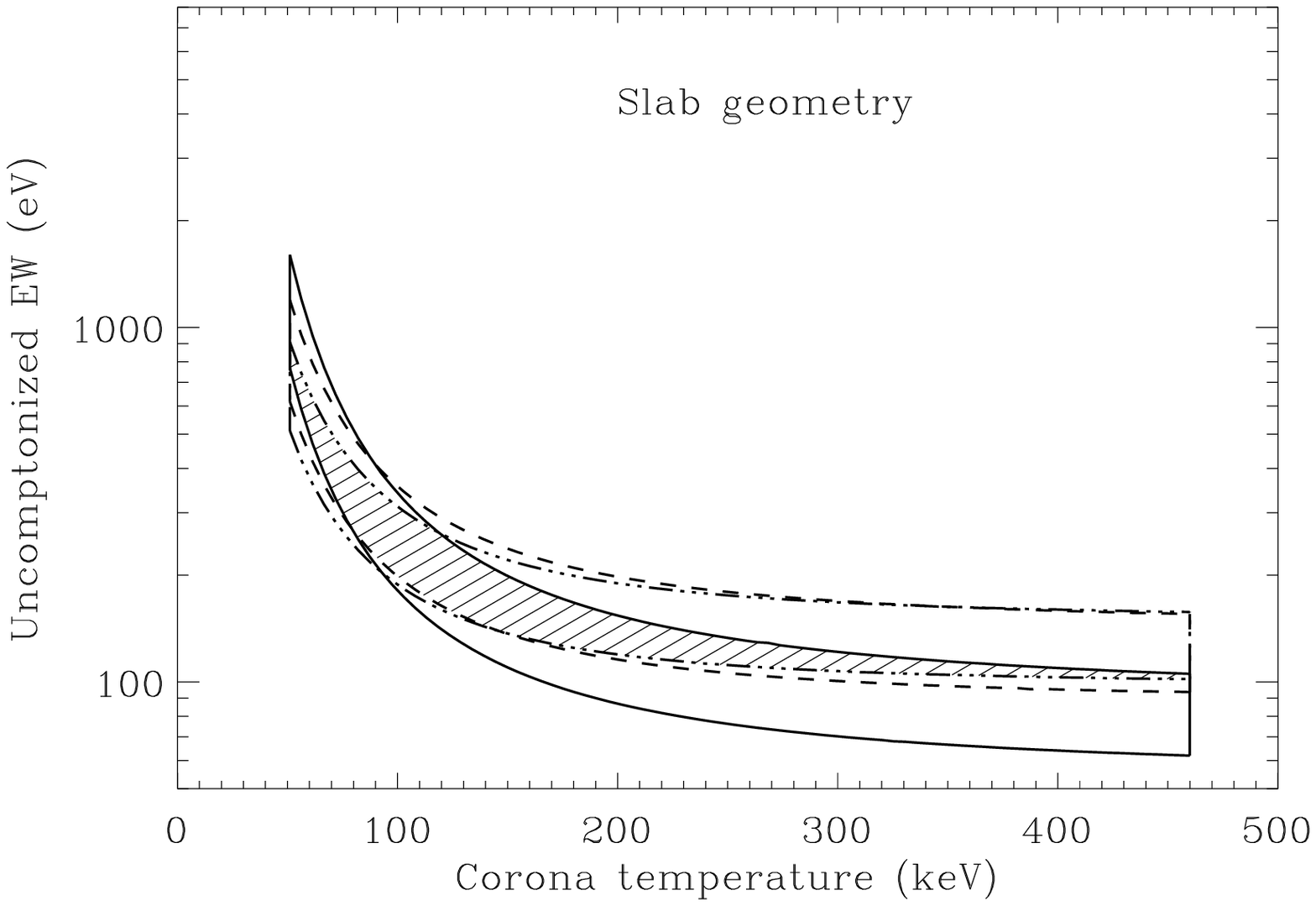}&
\psfig{angle=0,width=0.45\textwidth,file=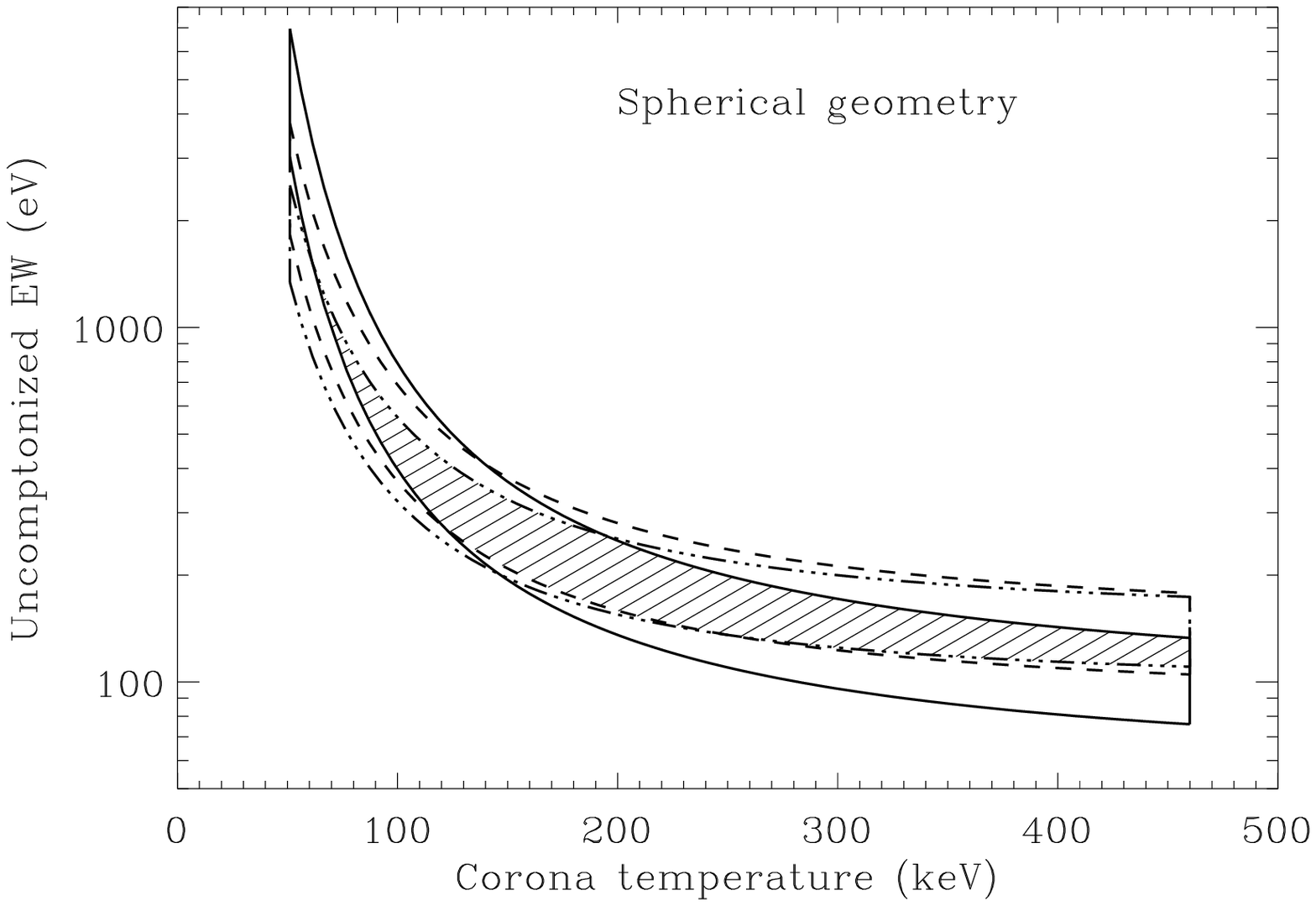}
\end{tabular}
\caption{The solid, dashed and dotted-dashed lines are the 90\% envelopes
of the uncomptonized iron line EW computed for the three spectral indexes
reported by LZ00, i.e. $\Gamma=$1.62, 1.82 and 1.94 respectively. The
hatched area correspond to the intersection of the three envelopes. For
any fixed value of $EW_{\rm dir}\ \gta$ 100 eV, this area delimits the
range of acceptable common values of the coronal temperature in the three
samples.  Alternatively, for any fixed value of $kT_e$, the hatched area
corresponds to the acceptable range for the broad line equivalent widths.
The left and right plots show the case of a slab and spherical geometry
respectively.}
\label{EWcorr}
\end{figure*}
In this case, and following R00, we can also set an upper limit on the
equivalent width of the uncomptonized broad iron line consistent with its
non detection by ASCA.  R00 have estimated that the corresponding
face--on EW { of a possible broad line} has to be smaller than $\sim$
200--300 eV { to be consistent with the data}. This estimate takes
into account the limb darkening and relativistic bending effects but not
the Comptonization of the iron line photons in the corona. If such an
effect is also taken into account (and still assuming an inclination of
82$^{\circ}$), { the limits are less strict} and a broad line with a
face--on uncomptonized EW$\sim$0.8--1.2 keV (for $\tau\sim$0.05--0.1)
would still be consistent with the ASCA data. { Then if there were a
broad line present at the level seen in most other AGN ($\sim$100--300
eV, Nandra et al. 1997; Lubinski \& Zdziarski 2000), we do not expect to
see it. The data thus do not require that the inner part of the
accretion disc be truncated as suggested in R00.}\\

% Such a large broad line EW (well above the mean one observed by Nandra
%et al. 1997) seems rather unlikely.  Therefore, given the high
%inclination angle of this source, the non-detection of any intrinsically
%broad iron line is naturally explained by the presence of a hot
%comptonizing corona that surrounds the line emitting region, and does not
%require that the inner part of the accretion disc be truncated as
%suggested in R00.\\

%%%%%%%%%%%%%%%%%%%%%%%%%%%%%%%%%%%%%%%%%%%%%%%%%%%%%%%%%%%%%%%%%%%%%%%%%%%%
\subsection{Correlations between reflection features and X-ray slope}
%%%%%%%%%%%%%%%%%%%%%%%%%%%%%%%%%%%%%%%%%%%%%%%%%%%%%%%%%%%%%%%%%%%%%%%%%%%%
The Comptonization effects we have discussed so far may influence the
apparent correlation we observe between the reflection features
characteristics (broad Fe line EW, reflection normalization) and the
other parameters of the high energy continuum.\\

In a recent paper LZ00 have analyzed the complete available ASCA database
to re--examine the issue of the strength and width of the Fe K$\alpha$
line in Seyfert galaxies. An interesting results they report is an
apparent correlation between the broad iron line EW and the spectral
index in this class of objects. The correlation is actually observed in
the average spectra of Seyferts galaxies grouped according to their
spectral hardness. As underlined by the authors, this correlation may be
naturally explained in the framework of thermal Comptonization
model. Indeed, in this case, the power law slope of the
X--ray/$\gamma$--ray spectrum emitted by the hot comptonizing corona is
related to the rate of cooling by incident soft photons. Then the
spectral slope will be correlated with the broad line EW, provided that
the main source of the cooling photons is emitted by the same medium that
is responsible for the observed reflection features. Furthermore, this
interpretation is consistent with the $\Gamma$--$R$ correlation observed
by Zdziarski et al. \shortcite {zdz99} in a sample of Seyfert and
galactic black hole candidates (see also Gilfanov et al. 1999; Matt
2000).\\
%A specific model with a mildly relativistic bulk motion of the
%hot plasma above an accretion disc (Beloborodov 1999a) can then
%quantitatively explain the correlation (Zdziarski et al. \shortcite {zdz99}).\\

{ We show here that the Comptonization effects in coronae of different
intrinsic properties (i.e. different temperature and optical depth)
between different objects may also contribute to produce this kind of
correlation.}\\
%We show here that this correlation may also be explained by the
%Comptonization of broad iron lines in coronae of different intrinsic
%properties (i.e. different temperature and optical depth).\\ 
In thermal Comptonization theory, $\tau$ and $kT_{\rm e}$ are related to
the X--ray spectra photon index $\Gamma$ through the approximate relation
\cite{ost74,poz76,zdz85}:
\begin{equation}
\Gamma=-\frac{\ln{P_{\rm sc}}}{\ln{A}}.
\label{Psc}
\end{equation}
In this equation, $P_{\rm sc}$ is the average scattering probability, its
expression depending on the coronal geometry, and $\displaystyle
A=1+4\frac{kT_{\rm e}}{m_{\rm e}c^2}+16\left(\frac{kT_{\rm e}}{m_{\rm
e}c^2}\right)^2$ is the average photon energy amplification per
scattering. In the following, we only treat the cases of slab and
spherical geometry. The expression of $P_{\rm sc}$ in these cases can be
found in Zdziarski et al. \shortcite {zdz94} and Wardzinski \& Zdziarski
\shortcite {war00} respectively.\\

For a given value of the photon index, we can then use Eq. (\ref{Psc}) to
compute the optical depth $\tau$ for different values of $kT_{\rm
e}$. Assuming that the broad iron lines observed by LZ00 have been
comptonized in the corona, we can then compute the EW of these lines
before Comptonization, $EW_{\rm dir}$, by inverting Eq. \ref{EWeq} (we
fix the inclination angle to 45 degrees, which is the best fit value
obtained by LZ00). In Fig.  \ref{EWcorr}a and \ref{EWcorr}b
(corresponding to a slab and a spherical geometry, respectively) we have
plotted the 90\% confidence envelopes for $EW_{\rm dir}$ and $kT_{\rm e}$
obtained for each of the three average values of the spectral indexes
reported by LZ00, i.e. $\Gamma$= 1.62, 1.82 and 1.94.
%, as functions of the coronal temperature.
%, which is assumed to be the same for the three classes.
These envelopes have been computed by taking into account the 90\% errors
of the photon index and of the observed EW reported by LZ00. The solid,
dashed and dotted-dashed envelopes in these figures correspond to
$\Gamma$=1.62, 1.82 and 1.94, respectively. The hatched surfaces
correspond to the region of the parameter space where the $EW_{\rm dir}$
computed for the different photon index classes are equal within the
errors.\\

Interestingly, the values of $EW_{\rm dir}$ are consistent with each
other (within the errors) for a relatively large range of coronal
temperatures and optical depths, particularly for the spherical geometry
where $EW_{\rm dir}\simeq$ 100--150 eV for $kT_{\rm e}\ \gta$ 150 keV and
$\tau\ \lta$ 0.6. In the slab case, $EW_{\rm dir}\simeq$ 100--150 eV for
100 $\lta\ kT_{\rm e}\ \lta$ 250 keV and 0.6 $\gta\ \tau\ \gta$ 0.1.
%These results thus suggest that the $EW$--$\Gamma$ correlation observed
%by LZ00 may simply result from the Comptonization of intrinsically broad
%iron lines, with similar EW (before Comptonization) in different objects,
%but in hot coronae having different physical characteristics
%(i.e. different $\tau$ and $kT_{\rm e}$).\\
{ These results thus show that the presence in different objects, all
having broad lines with similar EW (before Comptonization), of hot
coronae with different physical characteristics (i.e. different $\tau$
and $kT_{\rm e}$), may result in an apparent $EW$--$\Gamma$ correlation
(as the one observed by LZ00).\\

It is worth noting that the main implication of this interpretation is
that the coronae producing the harder spectra (smaller $\Gamma$) must
also have the larger optical depth. Such behaviour {\it necessarely goes
with a change of the Compton parameter} (i.e. a change of the geometrical
and/or energetical properties of the corona) since for constant $y$ a
correlation between $\tau$ and $\Gamma$ is expected (Haardt et
al. 1997).\\

We note that, in this picture, we also expect a correlation between the
observed reflection normalization $R$ and the photon index
$\Gamma$. Indeed, as shown in section \ref{simu}, for larger optical
depth the Comptonization effects smooth the reflection hump so that
fitting with an uncomptonized reflection gives smaller $R$. This goes in
the sense of the results of Z99, although, as shown in section 5, we do
not expect strong effects ($\lta$ 30\% for inclination angles smaller
than 60$^{\circ}$) on the reflection normalization.\\

A possible reason for an anticorrelation between $\tau$ and $\Gamma$ may
be the following. Suppose that the corona+disk configuration is the one
proposed by Z99 i.e. a central hot plasma + cold disc model, the inner
radius of the disk being able to vary. In this interpretation, the
increase of the inner radius of the disk, which will produce a decrease
of $R$ and $\Gamma$, may be due to the evaporation of part of the inner
regions of the disk (due to some disk instabilities, Meyer et al. 2000;
Menou et al. 2000; Turolla \& Dullemond 2000) in the corona, thus
resulting in a increase of the corona optical depth. Interestingly, in
this case, the comptonization effects would reduce the apparent
reflection component and thus could explain the small $R$ values observed
at small $\Gamma$, emphasizing the strength of the observed
correlation.\\

%Amusingly, if the picture above does not apply at all, e.g the $y$
%parameter is indeed constant between all the objects of the sample
%(i.e. $\tau$ and $\Gamma$ are correlated), the observed correlations
%would be dimed by the comptonization effects rather than emphasized,
%meaning that they may be intrinsically stronger than what we observed!\\

In conclusion, the main point of this section is that, due to likely
different corona characteristics from one object to the other, the
comptonization effects may have some influence to produce and/or
emphasize the observed correlations between the reflection features and
X-ray slope.}

%%%%%%%%%%%%%%%%%%%%%%%%%%%%%%%%%%%%%%%%%%%%%%%%%%%%%%%%%%%%%%%%%%%%%%%%%%%%
\subsection{Consequences for the X-ray background}
\begin{figure}
%\begin{figure}[t]
\psfig{angle=0,width=\columnwidth,file=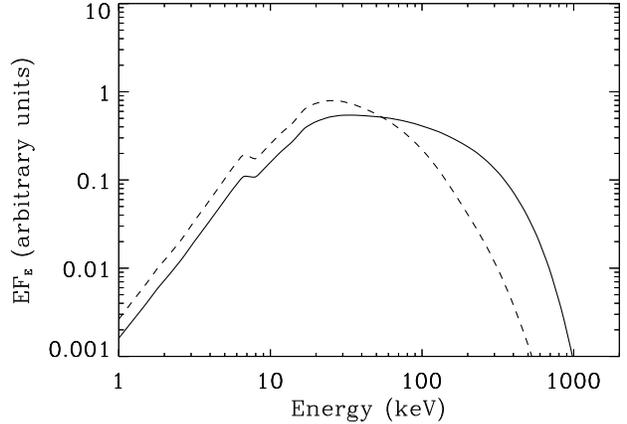}
\caption{An angle-averaged reflection hump calculated for $kT_{\rm e} =
90 keV$ and $\tau=0.35$ (corresponding to $y=0.6$). The solid line
represents the comptonized reflection, the dashed the direct one.  The
slight shift of the peak energy of the hump is an indication of the
predicted effect of the Comptonizing corona on the location of the peak
energy in the X-ray background.}
\label{angle_avg}
\end{figure}

%%%%%%%%%%%%%%%%%%%%%%%%%%%%%%%%%%%%%%%%%%%%%%%%%%%%%%%%%%%%%%%%%%%%%%%%%%%%
After the most recent surveys (Mushotzky et al. 2000; Hasinger et
al. 2001) there is now a tantalizing evidence that the X-ray Background
(XRB) is mainly due to the integrated emission of single sources,
i.e. AGNs lying at cosmological distances.  All the spectral models for
the XRB assume that the intrinsic AGN spectrum (before any absorption in
a putative large scale molecular torus) is the sum of a power-law
continuum and an uncomptonized reflection.  We expect that the effects
described here can be of some relevance for any detailed fitting model
for the XRB.  In Fig. \ref{angle_avg} we show, as an illustrative case,
an angle averaged reflection hump calculated for $kT_{\rm e} = 90 keV$
and $\tau=0.35$ (corresponding to $y=0.6$) for both the comptonized
(solid line) and uncomptonized (dashed line) case.  Indeed, in all
current models \cite{mad94,wil99,gil01}, for an average continuum slope
of about 2, the peak of the XRB (in $\nu F_\nu$) is associated with the
peak of the reflection hump.  Furthermore, there is evidence that the
observed peak of the X-ray background is located at slightly higher
energies than those predicted by the standard models that neglect the
effect of Comptonization in the corona.\\

Although many uncertainties come into the exact determination of the
possible consequences of such effect on the XRB shape, as the temperature
and optical depth distribution in the different sources or the total
coronal covering fraction, the work we have presented here should
indicate that Comptonization of the reflection component in the corona
have to be taken into account when dealing with accurate spectral fitting
of the X-ray background.

%%%%%%%%%%%%%%%%%%%%%%%%%%%%%%%%%%%%%%%%%%%%%%%%%%%%%%%%%%%%%%%%%%%%%%%%%%%%
%\subsection{Discussion}
%%%%%%%%%%%%%%%%%%%%%%%%%%%%%%%%%%%%%%%%%%%%%%%%%%%%%%%%%%%%%%%%%%%%%%%%%%%%
%%%%%%%%%%%%%%%%%%%%%%%%%%%%%%%%%%%%%%%%%%%%%%%%%%%%%%%%%%%%%%%%%%%%%%%%%%%%
\section{Conclusion}
%%%%%%%%%%%%%%%%%%%%%%%%%%%%%%%%%%%%%%%%%%%%%%%%%%%%%%%%%%%%%%%%%%%%%%%%%%%%
In this paper, we have shown that the effects of a comptonizing corona on
the appearance of the reflection components can be relatively important,
modifying the shape of the reflection hump and the iron line equivalent
width measurements. We have studied in detail the dependence of these
effects on the physical (i.e. the temperature $kT_e$ and optical depth
$\tau$) and geometrical (i.e. the inclination angle) parameters of the
corona, mainly focusing on the case of a slab geometry. { This is an
extreme case since the corona covers all the reflecting material and
consequently the comptonization effects are large}. The results of this
study can be summarized as follows:
\begin{itemize}
\item Due to the smoothing and shifting towards high energies of the
comptonized reflection hump, the main effects on the emerging spectra
appear at energies below and above $\sim kT_{\rm e}$.
\item They are larger for larger optical depth of the corona and/or
larger inclination angle of the corona--disk configuration.
\item Fitting Comptonization models taking into account comptonized
reflection by the usual cut--off power law + uncomptonized reflection
models, leads to an underestimation of the reflection normalization and
an overestimation of the high energy cut--off.
\item The Comptonization effects may strongly reduce the equivalent width
of the observed iron line, especially at large inclination angles.
\end{itemize}

These effects may have important consequences on the physical
interpretation of the presence and/or absence of reflection features in
astrophysical objects. As an example, we have studied the case of the
galaxy NGC 4258. With the assumption that the X--ray emitting region has
a disk--corona configuration and given the high inclination angle of the
accretion disk in this source, { the comptonization effects enable to find
less strict limits on the EW of a possible broad Iron line, then
explaining its non-detection}. Besides, in this picture, the narrow iron
line observed in this object is believed to originate preferentially from
matter not associated with the accretion disk since, in the contrary
case, it would require a primary (i.e. before Comptonization) iron line
with a (unlikely) too large EW.\\

{ We also find that the presence of a comptonizing corona can produce
and/or emphasize the correlations between the reflection
features (like the iron line equivalent width or the covering fraction)
and the X--ray spectral index.
%We note however that the main implication of
%this interpretation is that the coronae producing the harder spectra must
%also have the larger optical depth. This necessarely means a change of
%the Compton parameter (i.e. a change of the geometrical and/or
%energetical characteristic of the corona) between the different objects
%of the sample.  In any case, and 
Then, similarly to the effects produced by a bad modelisation of the
complex ionization pattern expected at the surface of the
X-ray--irradiated reflecting material, these different effects are to be
properly taken into account to correctly interpretate these
correlations.}\\

Finally, we underline the possible importance of the Comptonization
effects on the reflection shape when dealing with accurate spectral
fitting of the X-ray background. Indeed, there is evidence that the
observed peak of the X-ray background is located at slightly higher
energies than those predicted by the standard models that neglect the
effect of Comptonization in the corona.\\

%\noindent
\section*{Acknowledgments}
%{\sl Acknowledgements:} 
We are grateful to the referee, Chris Done, for the careful revision of
the paper and her valuable comments and suggestions.  POP acknowledges a
grant of the European Commission under contract number ERBFMRX-CT98-0195
(TMR network "Accretion onto black holes, compact stars and
protostars"). We thank J. Malzac for useful suggestions. We also thank
F. Fiore for having given to us the draft on NGC 4258 in advance of
publication.

%%%%%%%%%%%%%%%%%%%%%%%%%%%%%%%%%%%%%%%%%%%%%%%%%%%%%%%%%%%%%%%%%%%%%%%%%%%%%%

\label{lastpage}

\end{document}